\documentclass[12pt]{article}
\usepackage{amssymb,amsmath}
\usepackage[affil-it]{authblk}
\usepackage{graphicx}
\usepackage{epstopdf}
\usepackage[top=0.75in, bottom=0.75in, left=0.75in, right=0.75in, dvips]{geometry}
\usepackage{caption}
\begin{document}
\textwidth 10.0in
\textheight 9.0in
\topmargin -0.60in
\title{Renormalization Scheme Dependence  \\
 and the Renormalization Group Beta Function}
\author[1,2]{F.A. Chishtie\thanks{fchishti@uwo.ca}}
\author[3,4]{D.G.C. McKeon \thanks{dgmckeo2@uwo.ca}}
\affil[1] {Department of Physics and Astronomy, University of Western Ontario, London, ON N6A 3K7, Canada}
\affil[2] {Theoretical Research Institute, Pakistan Academy of Sciences (TRIPAS) Islamabad, Pakistan 44000}
\affil[3] {Department of Applied Mathematics, University of Western Ontario, London, ON N6A 5B7, Canada}
\affil[4]{Department of Mathematics and Computer Science, Algoma University, Sault St.Marie, ON P6A 2G4, Canada}
   
\maketitle
  
\maketitle

\noindent

\begin{abstract}
The renormalization that relates a coupling ``a" associated with a distinct renormalization group beta function in a given theory is considered.  Dimensional regularization and mass independent renormalization schemes are used in this discussion.  It is shown how the renormalization $a^*=a+x_2a^2$ is related to a change in the mass scale $\mu$ that is induced by renormalization. It is argued that the infrared fixed point is to be a determined in a renormalization scheme in which the series expansion for a physical quantity $R$ terminates.
\end{abstract}

\section{Introduction}

Elimination of divergences arising in the computation of radiative effects in quantum field theory results in the introduction of an unphysical mass scale parameter $\mu$.  Explicit dependence on $\mu$ must be offset by implicit dependence on $\mu$ through the parameters (couplings, masses and field strengths) characterizing the theory,  resulting the renormalization group (RG) equations [1-3].  If dimensional regularization [4-6] is used in conjunction with mass independent renormalization [7, 8], it is possible to make additional finite renormalizations that lead to further RG equations.

We will examine the effect of finite renormalizations in the QCD calculation of the process $e^+e^- \rightarrow$ hadrons with cross section $R_{e^+e^-}$.  The strong fine structure constant is $\alpha = a\pi$ and all quarks are taken to be massless.  A variety of renormalization schemes (RS) will be considered; minimal subtraction (MS) [8], a scheme due to 't Hooft in which the RG function $\beta$ associated with $a$ has no contribution beyond second order [9, 10, 11], and a scheme in which no radiative corrections beyond second order contribute to the expansion of $R_{e^+e^-}$ in powers of $a$ [12, 13].  The latter two schemes only involve RS invariant quantities. In each of these schemes $a = 0$ is an ultraviolet fixed (UV) points; that is, as the centre of mass energy scale $Q$ increases, the couplant $a$ goes to zero (``asymptotic freedom'') [8, 14-17].  We will consider the possibility of there also being an infrared (IR) fixed point, in which $a$ goes to some finite value as $Q$ decreases to zero. The scheme in which the expansion of $R_{e^+e^-}$ in powers of $a$ is finite is argued to be the only RS to be of relevance, as in this scheme the behaviour of the infinite series in powers of $a$ that occurs in other schemes is not a problem. 
 
\section{Finite Renormalization}

The cross section $R_{e^+e^-}$ can be expressed as a power series in the couplant $a$
\begin{equation}
R_{e^+e^-} = 3\left( \sum_i q^2_i\right) (1 + R)
\end{equation}
where the $n$ loop contribution to $R$ in perturbation theory is given by the term of order $a^{n+1}$ in the expansion
\begin{equation}
R = \sum_{n=0}^\infty \sum_{m=0}^n  T_{n,m} a^{n+1} L^m \quad  
\left( T_{0,0} = 1, \; L = b \ln \frac{\mu}{Q} \right).
\end{equation}
The explicit dependence of $R$ on $\mu$ through $L$ is cancelled by its implicit dependence through $a\left( \ln \frac{\mu}{\Lambda}\right)$ where
\begin{align}
\mu \frac{da}{d\mu} &= -ba^2 (1 + ca + c_2 a^2 + \ldots )\\
&\equiv \beta (a). \nonumber
\end{align}
The solution to eq. (3) is taken to be [18]
\begin{equation}
\ln \left( \frac{\mu}{\Lambda}\right) = \int_0^{a(\ln\frac{\mu}{\Lambda})} \frac{dx}{\beta(x)} + \int_0^\infty \frac{dx}{bx^2(1+cx)}.
\end{equation}

The general relation between the bare couplant $a_B$ appearing in the initial QCD Lagrangian and the renormalized couplant $a$ when using a mass independent renormalization scheme with dimensional regularization is [8]
\begin{equation}
a_B = \mu^{-\epsilon} \left[ A_0(a) + \frac{A_1(a)}{\epsilon} + \frac{A_2(a)}{\epsilon^2} + \ldots \right]
\end{equation}
where $\epsilon = -4 + n$, $n$ being the number of dimensions.  Since $a_B$ is independent of $\mu$
\begin{equation}
\mu \frac{da_B}{d\mu} = 0 = \left( \mu \frac{\partial}{\partial \mu} +\beta (a) \frac{\partial}{\partial a}\right) \mu^{-\epsilon} \left[ A_0 + \frac{A_1}{\epsilon} + \frac{A_2}{\epsilon^2} + \ldots \right].
\end{equation}
Eq. (6) is satisfied at orders $\epsilon$, $\epsilon^0$ provided
\begin{align}
\beta (a) &= \frac{1}{A_0^\prime} \left( A_1 -  \frac{A_0}{A_0^\prime} A_1^\prime \right) + \frac{A_0}{A_0^\prime} \epsilon \nonumber \\
&  = - \left( \frac{A_0}{A_0^\prime}\right)^2 \frac{d}{da} \left( \frac{A_1}{A_0}\right) + \left( \frac{A_0}{A_0^\prime}\right)\epsilon \;.
\end{align}
Terms of order $\epsilon^{-n}$ $(n = 1,2 \ldots)$ in eq. (6) fix $A_2, A_3 \ldots$ in terms of $A_0$ and $A_1$.

The function $A_0$ in eq. (5) is not fixed; the MS RS corresponds to selecting $A_0 (a) = a$.  Using $\overline{a}$ to denote the MS fine structure constant, then by eq. (7)
\begin{equation}
\overline{\beta}(\overline{a}) = \left( \overline{A}_1 (\overline{a}) - \overline{a}\; \overline{A}^\prime_1 (\overline{a}) \right) + \overline{a}\epsilon
\end{equation}
is the MS $\beta$-function. We now can expand a general function $A_0$ as
\begin{equation}
A_0 (a) = a + x_2 a^2 + x_3 a^3 + \ldots \; ;
\end{equation}
the identification
\begin{equation}
\overline{a} = A_0 (a)
\end{equation}
constitutes a finite renormalization of $\overline{a}$.

In general, if we have two different couplings $a$ and $a^*$ such that they are related by the renormalization
\begin{align}
a^* &= a + y_2 a^2 + y_3 a^3 + \ldots \\
&\equiv \rho (a) \nonumber
\end{align}
then from the relation
\begin{equation}
\mu \frac{da^*}{d\mu} = \mu \frac{da}{d\mu} \frac{da^*}{da} \Longrightarrow
\beta^* (a^*) = \frac{d\rho (a)}{da} \beta (a)
\end{equation}
or by eqs. (3,11)
\begin{equation}\tag{13}
-b^*a^{*^{2}} \left( 1 + c^* a^* + c_2^* a^{*^{2}} + \ldots\right) = -ba^2 (1 + ca + c_2a^2 + \ldots) (1 + 2y_2a + 3y_3 a^2 + \ldots )
\end{equation}
we see that [19]
\begin{equation}\tag{14a}
b^* = b 
\end{equation}
\begin{equation}\tag{14b}
c^* = c 
\end{equation}
\begin{equation}\tag{14c}
c^*_2 = c_2 - cy_2 + y_3 - y_2^2 
\end{equation}
\begin{equation}\tag{14d} 
c_3^* = c_3 - 3cy_2^2  + 2(c_2 - 2c_2^*) y_2 + 2y_4 - 2y_2y_3
\end{equation}
\begin{align}\tag{14c}
c_4^* & = c_4 - 2y_4y_2 - y_3^2  + c\left(y_4 - y_2^3 - 6y_2y_3\right) + 3y_3 c_2 - 4y_3c_2^* \\
&\hspace{1cm}-6y_2^2 c_2^* + 2y_2c_3 -5y_2 c_3^* + 3y_5 \nonumber
\end{align}
etc .

Consequently, $b$ and $c$ are RS invariants while $c_2, c_3 \ldots$ are RS dependent.  The RG function $\beta$ is thus not unique; in addition to the one associated with MS in eq. (8), there is the 't Hooft scheme in which $c_n = 0\; (n \geq 2)$ so that $\beta$ consists of two terms 
\begin{equation}\tag{15}
\beta (a) = - ba^2 (1 + ca)
\end{equation}
or the particular $\beta$-function associated with $N = 1$ supersymmetric gauge theory [20, 21]
\begin{equation}\tag{16}
\beta (a) = \frac{-ba^2}{1-ca}.
\end{equation}

In principle, the result of eq. (16) could be altered upon making a finite renormalization of a. It was noted in ref. [18] that a RS can be characterized by the RS dependent coefficients $c_2, c_3 \ldots$ in eq. (3).  From eq. (13) it would appear that these coefficients could be identified with $y_3, y_4 \ldots$ appearing in $\rho (a)$  in eq. (11); in ref. [18] it was suggested that $y_2$ should be associated with the scale parameter $\mu$.  (In ref. [19] this coefficient is taken to be arbitrary and not related to $\mu$; dependence of $a$ on $y_2$ is analyzed as if $y_2$ were a free parameter independent on $\mu$.)  Identifying $y_2$ with $\mu$ is reasonable, as by eqs. (3, 12)
\begin{equation}\tag{17}
\frac{da}{\beta(a)} = \frac{da^*}{\beta^*(a^*)}
\end{equation}
and so by eq. (4)
\begin{equation}\tag{18}
\int_0^{a\left(\ln \frac{\mu}{\Lambda}\right)} \frac{dx}{\beta(x)} - 
\int_0^{a^*(\ln \frac{\mu^*}{\Lambda})} \frac{dx}{\beta^*(x)} =
\ln \left(\frac{\mu}{\mu^*}\right). 
\end{equation}
However, if $\mu = \mu^*$ this extra arbitrariness disappears and $y_2$ should vanish. 

To see this connection between $\mu$ and $y_2$ more explicitly, let us consider
\begin{equation}\tag{19}
\frac{da}{dc_i} = \beta_i(a) \qquad (i \geq 2);
\end{equation}
from
\begin{equation}\tag{20}
\mu \frac{\partial^2a}{\partial\mu \partial c_i} - 
\mu \frac{\partial^2a}{ \partial c_i\partial\mu} = 0
\end{equation}
it follows that [18]
\begin{align}\tag{21}
\beta_i (a) & = -b\beta (a) \int_0^a dx \frac{x^{i+2}}{\beta^2(x)}\\
 & \hspace{1cm}\approx \frac{a^{i+1}}{i-1}  \Bigg[ 1 + \left( \frac{(-i+2)c}{i}\right) a + 
\left( \frac{(i^2 -3i+2)c^2 + (-i^2+3i)c_2}{(i+1)i}\right)a^2 \nonumber \\
& \hspace{.5cm}+\left( \frac{(-i^3 + 3i^2 + 4i)c_3 + (2i^3-6i^2 + 4)cc_2 + (-i^3 + 3i^2 - 2i)c^3}{(i+2)(i+1)i}\right)a^3  + \ldots \Bigg].\nonumber 
\end{align}
(RS dependency in theories with renormalized masses is considered in refs. [22, 23].) If now $a$ and $a^*$ at the same value of $\mu$ are expanded as
\begin{equation}\tag{22}
a^* = a + \lambda_2 (c_i^*, c_i) a^2 + \lambda_3 (c_i^*, c_i) a^3 + \ldots
\end{equation}
then the equation
\begin{equation}\tag{23}
\frac{da^*}{dc_i} = 0 = \left( \frac{\partial}{\partial c_i} + \beta_i \frac{\partial}{\partial a}\right) \left( a + \lambda_2 a^2 + \ldots \right) 
\end{equation}
with the boundary condition $\lambda_n(c_i, c_i) = 0$ can be used to show that 
[12, 13  ] 
\begin{align}\tag{24}
a^* &= a+ \left(c_2^* - c_2\right) a^3 + \frac{1}{2} \left(c_3^* - c_3\right)
a^4 + \big[ \frac{1}{6} \left(c_2^{*2} - c_2^2\right)\\
&+ \frac{3}{2} \left(c_2^* - c_2\right)^2 - \frac{c}{6} \left(c_3^* -c_3\right) + \frac{1}{3} \left(c_4^* - c_4\right)\big] a^5 +\ldots  \nonumber
\end{align}
Eqs. (11,14) are consistent with eq. (24) only if $y_2 = 0$.

If now from eq. (14b) we see that
\begin{equation}\tag{25a}
y_3 = c_2^* - c_2 + cy_2 + y_2^2 
\end{equation}
so that from eqs. (14d, 25a) we obtain
\begin{equation}\tag{25b}
y_4 = \frac{1}{2} \left[ c_3^* - c_3 + \left(6c_2^* - 4c_2\right) y_2 + 5cy_2^2 + 2y_2^3\right];
\end{equation}
eqs. (14e, 25a, 25b) now lead to
\begin{align}\tag{25c}
y_5 &= \frac{1}{3}\Big\{\left( c_4^* - c_4\right)+ y_2\left(5c_3^* - 2c_3\right) + \left(4c_2^* - 3c_2+6y_2c\right)\left(c_2^* - c_2+cy_2+y_2^2\right)\nonumber \\
 &\hspace{1cm}+\left(c_2^* - c_2+cy_2+y_2^2\right)^2+ 6y_2^2c_2^* +y_2^3c+ \left( 2y_2 - c\right) \left[\frac{1}{2}\left( c_3^* - c_3\right) +  y_2\left(3 c_2^* -2 c_2\right)+ \frac{5}{2}cy_2^2 +y_2^3 \right] \nonumber \Big\} 
\end{align}
etc. \\

We now take $a^*$ to have the same $\beta$ function as $a$, but evaluated with mass scale $\mu^*$ rather than $\mu$.  If now we expand $a^*$ in terms of $a$ [12, 13, 24] so that
\begin{equation}\tag{26}
a^* = a + \left(\sigma_{21} \ell \right) a^2 + \left(\sigma_{31}\ell + \sigma_{32} \ell^2 \right) a^3 + \ldots\qquad \left( \ell = b \ln \frac{\mu}{\mu^*}\right)
\end{equation}
then as
\begin{equation}\tag{27}
\mu \frac{da^*}{d\mu} = 0 = \left( \mu \frac{\partial}{\partial \mu} + \beta (a) \frac{\partial}{\partial a} \right) \left[ a + \left(\sigma_{21} \ell \right) a^2 + \left(\sigma_{31} \ell+ \sigma_{32} \ell^2 \right)a^3 + \ldots \right]
\end{equation}
we find that
\begin{align}\tag{28}
a^* &= a + (\ell) a^2 + \left( c\ell + \ell^2\right) a^3 + \left(c_2\ell + \frac{5}{2} c\ell^2 + \ell^3\right) a^4 \\
&\hspace{2cm}+ \left[ c_3 \ell + \left( 3c_2 + \frac{3}{2} c^2\right) \ell^2 + \frac{13}{3} c\ell^3 + \ell^4\right] a^5 \nonumber \\
& \hspace{2cm}+\left[ c_4 \ell + \frac{7}{2}\left( c_3 + c_2 c\right)\ell^2 + \left( 6c_2 + \frac{35}{6} c^2\right) \ell^3 + \frac{77}{12} c\ell^4 + \ell^5\right] a^6 + \ldots \nonumber
\end{align}
This is identical to what is obtained from eqs. (11, 25) in the limit $c_i^* = c_i$ provided
\begin{equation}\tag{29}
y_2 = \ell.
\end{equation}
This is consistent with eq. (18) and with the observation in ref. [18] that 
$y_2$ is to be identified with the mass scale parameter $\mu$. It is also posible to use eqs. (24, 28) together to expand $a^*(\mu^*, c{_i}^*)$ in powers of $a(\mu, c{_i})$; the result is the same as eq. (11) with eqs. (25, 29). \

We now make the expansion
\begin{equation}\tag{30}
\overline{A}_1(\overline{a}) = \overline{\lambda}_2 \overline{a}^2 + 
\overline{\lambda}_3 \overline{a}^3 + \overline{\lambda}_4 \overline{a}^4 + \ldots
\end{equation}
where $\overline{A}_1(\overline{a})$ is associated with the MS RS.  We find from eq. (8) that in this scheme
\begin{equation}\tag{31a,b,c}
\overline{b} = \overline{\lambda}_2, \qquad \overline{c} = 2 \overline{\lambda}_3/ \overline{\lambda}_2, \qquad 
\overline{c}_n = (n+1) \overline{\lambda}_{n+2}/ \overline{\lambda}_2.
\end{equation}
If under a finite renormalization given by eq. (9) we end up with a coupling $a$ and a RG function $\beta(a)$ given by eq. (7), then the expansions of eqs. (3,9) show that if 
\begin{equation}\tag{32}
A_1(a) = \lambda_2 a^2 + \lambda_3 a^3 + \ldots
\end{equation}
then
\begin{equation}\tag{33a}
\lambda_2 = b
\end{equation}
\begin{equation}\tag{33b}
\lambda_3 = \frac{1}{2} b(c + 4x_2)
\end{equation}
\begin{equation}\tag{33c}
\lambda_4 = \frac{b}{3} \left( 7 x_3 + 2 x_2^2 + \frac{7}{2} cx_2 +c_2\right)
\end{equation}
\begin{equation}\tag{33d}
\lambda_5 = \frac{b}{4} \left( 10  x_4 + \frac{22}{3} x_2x_3 - \frac{4}{3} x_2^3 + \frac{5}{3} cx_2^2 + 6cx_3+ \frac{10}{3} x_3c_2 +c_3\right)
\end{equation}
etc.

Upon identifying $y_i$ with $x_i$, eq. (25) can be used to convert eqs. (33c, 33d) to
\begin{equation}\tag{34a}
\lambda_4 = b \left( \frac{7}{3}\overline{c}_2 + \frac{7}{2} x_2 c + 3x_2^2 -2c_2\right)
\end{equation}
\begin{equation}\tag{34b}
\lambda_5 = \frac{b}{4} \left[\frac{112}{3} \overline{c}_2x_2-24x_2c_2+ 40cx_2^2 + 16 x_2^3+ 6 c^2x_2+ 6c(\overline{c}_2-c_2)  + 5 \overline{c}_3 - 4c_3\right]
\end{equation}
etc. \\
Together, eqs. (9,14, 31, 33) can be used to show that 
\begin{equation}\tag{35a}
\lambda_2 a^2 + \lambda_3 a^3 + \lambda_4 a^4 + \ldots = 
\overline{\lambda}_2\overline{a}^2 + \overline{\lambda}_3 \overline{a}^3 + \overline{\lambda}_4 \overline{a}^4 + \ldots
\end{equation}
so that
\begin{equation}\tag{35b}
A_1 (a) = \overline{A}_1 (\overline{a}).
\end{equation}
We thus see that with $\beta(a)$ and $\overline{\beta}(\overline{a})$ being given by eqs. (7) and (8) respectively, eqs. (10) and (14) are consistent.

We now can examine the presence of an IR fixed point in light of finite renormalizations.

\section{Infrared Fixed Points}

Eq. (12) implies that if $\beta(a) = 0$, then $\beta^* (a^*) = 0$ as well, if $a^* = \rho (a)$.  In addition, from eq. (12) it also follows that
\begin{equation}\tag{36}
\frac{d\beta^*(a^*)}{da^*} = \left( \frac{d^2\rho(a)}{da^2} \beta(a) + 
\frac{d\rho(a)}{da}\frac{d\beta(a)}{da} \right) \Big/ \frac{d\rho(a)}{da}
\end{equation}
and so if $\beta(a) = 0$, then [25, 26]
\begin{equation}\tag{37}
\frac{d\beta^*(a^*)}{da^*} = \frac{d\beta(a)}{da}  .
\end{equation}
These arguments rely on $\rho(a)$ being a well defined function when $\beta(a) = 0$; for ill behaved functions $\rho(a)$, it may turn out that $\beta^*(a^*)$ is non zero or that eq. (37) is not satisfied [27, 28].

In refs. [12, 13] the problem of RS dependence was considered in the context of the cross section $R_{e^+e^-}$ of eq. (1).  It was shown that by applying the RG equation
\begin{equation}\tag{38}
\left( \mu \frac{\partial}{\partial\mu} + \beta (a) \frac{\partial}{\partial a}\right) R = 0
\end{equation}
it is possible sum the logarithm in eq. (2) so that the explicit dependence of $R$ on $\mu$ (through $L$) and its implicit dependence (through $a(\mu)$) cancel, leaving us with
\begin{equation}\tag{39}
R = \sum_{n=0}^\infty T_n a^n \left( \ln \frac{Q}{\Lambda}\right) \qquad (T_n \equiv T_{n,0})
\end{equation}
where $Q$ is centre of mass energy and $\Lambda$ is a scale parameter introduced in eq. (4).

The behaviour of the sum in eq. (39) can be affected by three things.  There is first the behaviour of $a \left( \ln \frac{Q}{\Lambda}\right)$ as $Q$ itself evolves, secondly the behaviour of $T_n$ as $n$ becomes large, and thirdly the convergence behaviour of the infinite sum itself.  But $R$ is invariant under the finite renormalization of eq. (11); we will now examine how $R$ is affected by two particular renormalization schemes.  We will then consider their implications for the IR fixed point. Of course this IR fixed point cannot incorporate non-perturbative effects in QCD; our discussion is entirely in the context of the perturbative expansion of eq. (2). Non-perturbative effects in the low energy regime would necessarily involve the emergence of low energy Goldstone Bosons (pions).

Changes in RS lead to compensating changes in $T_n$ and $a$ so that
\begin{equation}\tag{40}
\frac{d}{dc_i} R = 0 = \left(\frac{\partial}{\partial c_i} + \beta_i (a) \frac{\partial}{\partial a}\right) \sum_{n=0}^\infty T_n  a^{n+1} .
\end{equation}
This leads [12, 13] to a set of nested equations for $T_n$ whose solutions are 
\begin{align}\tag{41a-f}
T_0 &= \tau_0 =1, \quad T_1 = \tau_1, \quad T_2 = -c_2 + \tau_2, \quad T_3 = -2c_2\tau_1 - \frac{1}{2} c_3 + \tau_3\\
T_4 &= -\frac{1}{3} c_4  - \frac{c_3}{2} \left( - \frac{1}{3} c + 2\tau_1\right) + \frac{4}{3} c_2^2 - 3c_2 \tau_2 + \tau_4\nonumber\\
T_5 &= \left[ \frac{1}{3} cc_2^2 + \frac{3}{2} c_2c_3 + \frac{11}{3} c_2^2 \tau_1 - 4c_2\tau_3\right] - \frac{1}{2} \left[ \frac{1}{6} c^2c_3 - \frac{2}{3} c_3c \tau_1 + 3 c_3 \tau_2\right] \nonumber \\
& \hspace{1cm} - \frac{1}{3} \left[ - \frac{1}{2} c_4 c + \frac{1}{2} c_4\tau_1\right] - \frac{1}{4} c_5 + \tau_5\nonumber
\end{align}
etc. \\
In eq. (41), the $\tau_n$ are constants of integration and hence are RS invariants. They can be determined by evaluating the $T_n$ and $b$, $c$, $c_n$ in one particular RS (such as MS) upon explicitly computing the relevant Feynman diagrams.

Two particular RS are of special interest. In the first scheme, the $c_i$ are selected so that $T_n = 0 (n \geq 2)$.  From eq. (41) this means that 
\begin{align}\tag{42a-c}
c_2 &= \tau_2 \\
c_3 &= 2(-2c_2\tau_1 + \tau_3)\nonumber \\
&= 2 (-2 \tau_1\tau_2 + \tau_3)\nonumber \\
c_4 &= - \frac{3}{2}c_3 \left( - \frac{1}{3} c + 2\tau_1\right) + 4c_2^2 - 9c_2\tau_2 + 3\tau_4\nonumber \\
&= c\left( \tau_3 - 2\tau_1\tau_2\right) + 12 \tau_1^2 \tau_2 - 6\tau_1 \tau_3 - 5 \tau_2^2 + 3\tau_4 \nonumber 
\end{align}
etc. \\
In the second scheme due to 't Hooft we set $c_n = 0 (n \geq 2)$ [9-11], so that 
\begin{equation}\tag{43}
T_n = \tau_n.
\end{equation}
In the first instance, the series in eq. (39) collapses down to two terms
\begin{equation}\tag{44a}
R_{(1)} = a_{(1)} + \tau_1 a_{(1)}^2 \left(\ln \frac{Q}{\Lambda}\right)
\end{equation}
while in the second case we have the infinite series 
\begin{equation}\tag{44b}
R_{(2)} = \sum_{n=0}^\infty  \tau_n a_{(2)}^{n+1} \left(\ln \frac{Q}{\Lambda}\right).
\end{equation}
In the first case $a_{(1)}$ ``runs'' according to eq. (4) with the $c_i$ of eq. (3) being given by eq. (42).  In the second case, $a_{(2)}$ runs according to the 't Hooft $\beta$-function of eq. (14).  In eq. (44a), there is no possible problem associated with there being a divergent series for $R_{(1)}$ or of having diverging behaviour for the coefficients of $a_{(1)}^n$; one need only discuss how $a_{(1)}\left(\ln \frac{Q}{\Lambda}\right)$ behaves as $Q$ varies.  On the other hand, in the infinite series for $a_{(2)}$, the behaviour of $a_{(2)}\left(\ln \frac{Q}{\Lambda}\right)$ is completely known as upon integrating eq. (4) we obtain the Lambert W function [29-32].  With $\beta_{(2)}(a_{(2)})$ given by eq. (14) it is evident that $a_{(2)}$ has an UV fixed point at $a_{(2)} = 0$ (if $b > 0$) and an IR fixed point at $a_{(2)} = \frac{-1}{c}$ (if $ c < 0$).  As one [14-17] and two [33, 34] loop calculations show that for an SU(N) gauge theory with $N_f$ flavours of quarks,
\begin{equation}\tag{45a,b}
b = \frac{33 - 2N_f}{6}, \quad c = \frac{153 - 19N_f}{2(33-2N_f)}
\end{equation}
which means that in order to have asymtotic freedom $(b < 0)$ and a positive IR fixed point  $(c < 0)$ we must have
\begin{equation}\tag{46}
8 \leq N_f \leq 16.
\end{equation}

However, even if $N_f$ satisfies the restrictions of eq. (46), the series of eq. (44b) could be badly behaved, or alternatively if eq. (46) is not satisfied, this series could be well behaved.  We are thus led to consider the finite sum of eq. (44a).  

The equality
\begin{equation}\tag{47}
R_{(1)} = R_{(2)}
\end{equation}
is consistent with eq. (24). However, since $R_{(2)}$ involves an infinite series whose convergence is not known, the IR behaviour of $a_{(2)}$ is not necessarily a reflection on the IR behaviour of $a_{(1)}$ since $R_{(1)}$ only involves a finite series in which convergence is no longer an issue. We contend that in analyzing the IR behaviour of $R_{e^+e^-}$ it is the behaviour of $a_{(1)}$ that is of relevance. We will now consider the IR behaviour of $a_{(1)}$ up to four loop order and thereby find out how $R_{(1)}$ behaves as ${Q\to 0}$.

Explicit calculation shows that with $N_f = 3$ active flavours of quarks [13],
\begin{equation}\tag{48a}
\hspace{-1.2cm}\tau_1 = 1.6401
\end{equation}
\begin{equation}\tag{48b}
\tau_2 = -5.812885185 
\end{equation}
\begin{equation}\tag{48c}
\tau_3 = -81.73499303 
\end{equation}
etc. \\
These values follow from the four-loop calculations of $T_n$ and $\beta(a)$ done in the $\overline{MS}$ scheme [35, 36].  With these values of $\tau_1$, $\tau_2$, $\tau_3$ it follows from eq. (42) that in the RS associated with $a_{(1)}$,
\begin{equation}\tag{49a}
\hspace{-.5cm}c_2 = -5.812885185
\end{equation}
\begin{equation}\tag{49b}
c_3 = -125.3351844092 
\end{equation}
etc. \\
Since with $N_f = 3$ [8, 14-17],
\begin{equation}\tag{50a}
b = 9/4
\end{equation}
and [33, 34]
\begin{equation}\tag{50b}
c = 16/9
\end{equation}
we see that the function $\beta_{(1)}(a_{(1)})$ is given by
\begin{equation}\tag{51}
\beta_{(1)}\left(a_{(1)}\right) = -2.25 a_{(1)}^2 \left( 1 + 1.77778 a_{(1)} - 5.812885185 a_{(1)}^2 - 125.3351844092 a_{(1)}^3 + \ldots \right).
\end{equation}

We now consider only the four-loop $\left(\mathcal{O}\left(a_{(1)}^5\right)\right)$ contribution to $\beta_{(1)}$ in eq. (51).  The function
\begin{equation}\tag{52}
f(x) = 1 + 1.77778x - 5.812885185x^2 - 125.3351844092x^3
\end{equation}
has only one positive zero, and that is found by Newton's Method to be at 
\begin{equation}\tag{53}
x \approx .20743211594.
\end{equation}

Consequently $\beta_{(1)}$ has an IR fixed point at 
\begin{equation}\tag{54}
a_{(1)} \approx .20743
\end{equation}
if we use only up to the four-loop contributions to $\beta_{(1)}$.  (In MS, $\overline{\beta}(\overline{a})$ has been computed to five-loop order [37], but to get 
$\beta_{(1)}(a_{(1)})$ to this order we would also require $R_{e^+e^-}$ to five-loop order; this is as yet unknown.)  From eqs. (1), (44a) and (54) it follows that
\begin{equation}\tag{55}
R_{e^+e^-}\Big/\left(3 \sum_i q_i^2\right) = 1 + (.20743) + (1.6401)(.20743)^2 \approx 1.278.
\end{equation}
This is the limit of $R_{e^+e^-}$ as $Q\rightarrow 0$ when we use $R_{(1)}$ in eq. (44a) and keep contributions only up to four-loop order.  This is consistent with $R_{e^+e^-}$ as presented in ref. [13].

Both $c_2$ and $c_3$ when using $\beta_{(1)}$ are negative (see eq. (49)) and so it is not unreasonable to anticipate that $c_4$ is also negative.  In fig. (1) we have plotted possible values of $c_4$ versus zero of $\beta_{(1)}$ to five loop order (using eqs. (49, 50) for $b$, $c$, $c_2$, $c_3$) when there are three active flavours of quarks.  We see that for $-2000 \leq c_4 \leq 300$, $\beta_{(1)}$ to five loop order has zeros lying between $.1$ and $.3$ which is quite reasonable.  We note that by eq. (41f), $\tau_5$ and $T_5$ both vary linearly with $c_4$.

If we use eq. (24) to relate $a_{(1)}$ to $a_{(2)}$ we see from eq. (42) that
\begin{equation}\tag{56}
a_{(2)} = a_{(1)} - \tau_2 a_{(1)}^3 + \left(2\tau_1\tau_2 - \tau_3\right) a_{(1)}^4 + \ldots \; .
\end{equation}
Using eq. (48), we find that the value of $a_{(2)}$ corresponding to $a_{(1)} = .20743$ is
\begin{equation}\tag{57}
a_{(2)} = .37533\; .
\end{equation}
This clearly is not an IR fixed point for the function $\beta_{(2)}(a_{(2)})$; with
$\beta_{(2)}$ given by eq. (15), the only value of $a_{(2)}$ for which $\beta_{(2)}$ vanishes is given by $a_{(2)} = - \frac{1}{c}$ which for $N_f = 3$ is, by eq. (50b), negative -- an unacceptable value.  However, by eq. (47), we see that as $Q \rightarrow 0$, $1 + R_{(1)} \rightarrow 1.278$, then $1 + R_{(2)}$ must also approach this value even though $a_{(2)}$ approaches a value given by eq. (57) which is not an IR fixed point of $\beta_{(2)}$.  Actually $a_{(2)}$ approaches this value only if we keep the three terms of eq. (24) given in eq. (56); if we were to simply consider integrating eq. (2) to find $a_{(2)}\left(\ln \frac{Q}{\Lambda}\right)$ exactly, it is evident that as $Q \rightarrow 0$, $a_{(2)} \rightarrow \infty$ if $b > 0, c > 0$.  This indicates that the full series of eq. (56) diverges as $Q \rightarrow 0$ (when $a_{(1)}$ approaches an exact IR fixed point and $a_{(2)}$ diverges).

We also note that with $N_f = 3$, the $\beta$ function in the $\overline{MS}$ scheme at four and five loop order has no positive roots, indicating that in the $\overline{MS}$ scheme, there is no IR fixed point, at least to this order in perturbation theory.  This is discussed in ref. [39].

An interesting third RS is one in which $c_2$ is allowed to vary while $c_i (i > 2)$ vanishes.  By eq. (41), in this scheme $R$ is given by
\begin{align}
R_{(3)} = a_{(3)} &+ \tau_1 a_{(3)}^2 + (-c_2 + \tau_2) a_{(3)}^3\nonumber \\
&\qquad
+ (-2 c_2 \tau_1 + \tau_3) a_{(3)}^4 + \left(\frac{4}{3} c_2^2 -3 c_2 \tau_2 + \tau_4\right)a_{(3)}^5 + \ldots \nonumber
\end{align}
while
\begin{equation}
\beta_3(a_{(3)}) = - b a_{(3)}^2 \left( 1 + c a_{(3)} + c_2 a_{(3)}^2\right).\nonumber 
\end{equation}

It appears to be possible to have a value of $c_2$ and an acceptable value of $a_{(3)}$ such that $\beta_{(3)} = 0$ and at order  $a_{(3)}^5$, $R_{(3)} = 1.278$; we find that $c_2 = -13.106$ and $a_{(3)} = .3523$.

We conclude that the perturbative expression for $1 + R_{(2)}$ appearing in eq. (44b) is an infinite series whose behaviour as $Q \rightarrow 0$ is such that it approaches a fixed value even though $a_{(2)}\left(\ln \frac{Q}{\Lambda}\right)$ is not an IR fixed point in this limit.  This fixed value for $1 + R_{(2)}$ is trivial to compute using eq. (44a) provided we can find an IR fixed point for $\beta_{(1)}$.  Upon only employing the first four terms of the expansion of eq. (51) for $\beta_{(1)}$, this IR fixed point is given by eq. (54).

\section{Discussion}

We have examined various aspects of RS dependency in perturbative QCD.  First of all, we have considered within the context of mass independent renormalization schemes how the parameters $y_i$ appearing in the finite renormalization of eq. (11) are related to the parameters $\mu$ and $c_i$ of eq. (3); we led to eqs. (25) (for $y_n$, $n > 2$) and eq. (29) (for $y_2$). Treating $y_2$ as an independent parameter as in ref. [19] should involve the mass scale parameter $\mu$. Secondly, the behaviour of QCD predictions for physical processes in the IR limit are considered.  After using RG summation to eliminate $\mu$ dependence, the RS scheme in which perturbative calculations give rise to just a finite number of contributions in powers of the coupling is taken to be the scheme that can be used to consider this IR limit.  This is because in this scheme one is not confronted with an infinite series in powers of the coupling whose behaviour is unknown. (Indeed, it has been argued that this series contains ``renormalons'' [9] and is at best asymptotic.)  The IR behaviour when using this finite series is controlled by the IR behaviour of $a_{(1)}$ in eq. (51) (for $N_f = 3$) if we only use results up to four-loop order.

We note that since the values of $\tau_i$ are determined by eq. (41) when $T_i$ and $c_i$ are computed in a particular renormalization scheme, it follows that the $\tau_i$ must be computed separately for distinct processes.  Thus, by eq. (42), the IR fixed point for $\beta_{(1)}$ is not the same in different processes.

We also would like to outline how the ``Principle of maximum conformality'' [38, 40] (PMC) or its improvement [41]  is related to the renormalization group summation employed in this paper.  Let us consider the sum of eq. (39) when using the RS of eq. (43) so that $R$ is given by eq. (44b).  It is now possible to expand $a_{(2)} \left(\ln \frac{Q}{\Lambda}\right)$ in terms of $a \left(\ln \frac{Q}{\Lambda}\right)$ where $a$ is the couplant in some other RS by using eq. (24)
\begin{equation}\tag{58}
a_{(2)}\left(\ln \frac{Q}{\Lambda}\right) = a\left(\ln \frac{Q}{\Lambda}\right) + (-c_2) a^3\left(\ln \frac{Q}{\Lambda}\right) + \frac{1}{2}(-c_3) a^4\left(\ln \frac{Q}{\Lambda}\right)
\end{equation}
\begin{equation}
\hspace{2cm}+ \left[ \frac{1}{6}(-c_2^2) + \frac{3}{2}(-c_2)^2 - \frac{c}{6}(-c_3)  + \frac{1}{3}(-c_4)\right] a^5\left(\ln \frac{Q}{\Lambda}\right) + \ldots \; .\nonumber
\end{equation}
Now in turn, $a\left(\ln \frac{Q}{\Lambda}\right)$ can be expanded in terms of $a\left(\ln \frac{\mu_n}{\Lambda}\right)$ using eq. (28) when $a_{(2)}^{n}\left(\ln \frac{Q}{\Lambda}\right)$ appears in eq. (44b).  This results in (with 
$a_n = a\left(\ln \frac{\mu_n}{\Lambda}\right)$ and  $\ell_n = b \ln \frac{\mu_n}{Q}$)
\begin{align}\tag{59}
R_{(2)} & = \tau_0 \left[ a_1 + \ell_1 a_1^2 + \left( c\ell_1 + \ell_1^2\right)a_1^3 + \left( c_2 \ell_1 + \frac{5}{2} c \ell_1^2 + \ell_1^3 \right) a_1^4 + \ldots \right]
\nonumber \\
& +\tau_1 \left[ a_2 + \ell_2 a_2^2 + \left( c\ell_2 + \ell_2^2\right)a_2^3 + \left( c_2 \ell_2 + \frac{5}{2} c \ell_2^2 + \ell_2^3 \right) a_2^4 + \ldots \right]^2\nonumber \\
& + \tau_2 \left[ a_3 + \ell_3 a_3^2 + \left( c\ell_3 + \ell_3^2\right)a_3^3 + \left( c_2 \ell_3 + \frac{5}{2} c \ell_3^2 + \ell_3^3 \right) a_3^4 + \ldots \right]^3\nonumber \\
& + \ldots . \nonumber
\end{align}
Upon grouping terms in eq. (59) in ascending powers of $a$ we obtain
\begin{align}\tag{60}
R_{(2)} & = \tau_0 a_1 + \left[ (\tau_0\ell_1) a_1^2 + \tau_1 a_2^2\right]\nonumber \\
& + \left\lbrace \left[ \tau_0 \left(c_1 + \ell_1^2\right)  a_1^3 + \tau_1 (2\ell_2) a_2^3 \right] + \tau_2 a_3^3 \right\rbrace \nonumber \\
& + \Big\{ \Big[  \tau_0 \left(c_2 \ell_1  + \frac{5}{2} c \ell_1^2 + \ell_1^3 \right)
 a_1^4 + \tau_1 \left( \ell_2^2 + 2 (c_2\ell_2 + \ell_2^2)\right) a_2^4 \nonumber \\
&+ \tau_2 (3\ell_3) a_3^4 \Big] + \tau_3 a_{4}^4 \Big\} + \ldots  .\nonumber
\end{align}

It is now possible to select $\mu_1, \mu_2, \mu_3 \ldots$ so that eq. (60) reduces to
\begin{equation}\tag{61}
R_{(2)} = \sum_{n=0}^\infty \tau_n a_{n+1}^{n+1};
\end{equation}
this entails solving
\begin{equation}\tag{62a}
\tau_0 \ell_1 = 0
\end{equation}
\begin{equation}\tag{62b}
\tau_0 (c_1 + \ell_1^2) a_1^3 + \tau_1 (2\ell_2) a_2^3 = 0
\end{equation}
\begin{equation}\tag{62c}
\tau_0 \left(c_2 + \ell_1 + \frac{5}{2}c \ell_1^2 + \ell_1^3\right) a_1^4 + \tau_1 \left(\ell^2_2 + 2 ( c_2 \ell_2 + \ell_2^2)\right) a_2^4 + \tau_2 (3\ell_3) a_3^4  = 0
\end{equation}
etc.\\
From eq. (62a), we see that $\mu_1 = Q$; solving for $\mu_2,\mu_3 \ldots$ etc. becomes progressively more difficult.  The resulting expression for $R_{(2)}$ in eq. (61) is now expressed in terms of scheme independent quantities $\tau_n$ and all dependence on $\mu$ has disappeared; $\mu_1, \mu_2 \ldots$ contain all of the explicit dependence on $c_2, c_3 \ldots$.  As a result, eq. (61) is equivalent to what is obtained using the approach to PMC used in ref. [38].

We wish to note that the low energy behaviour of $a(Q)$ has also been examined using light cone holography [40]. Our considerations have been limited to examining RS ambiguities that arise in using conventional perturbative evaluation of physical quantities in QCD.

In the future we hope to extend these considerations of perturbative expansions in quantum field theory to processes involving non-trivial masses and/or multiple couplings [23]. 

\section*{Acknowledgements}
A correspondence with A. Kataev and A. Deur is gratefully acknowledged and 
R. Macleod made a helpful comment.

\begin{figure}[hbt]
\begin{center}
\includegraphics[scale=0.9]{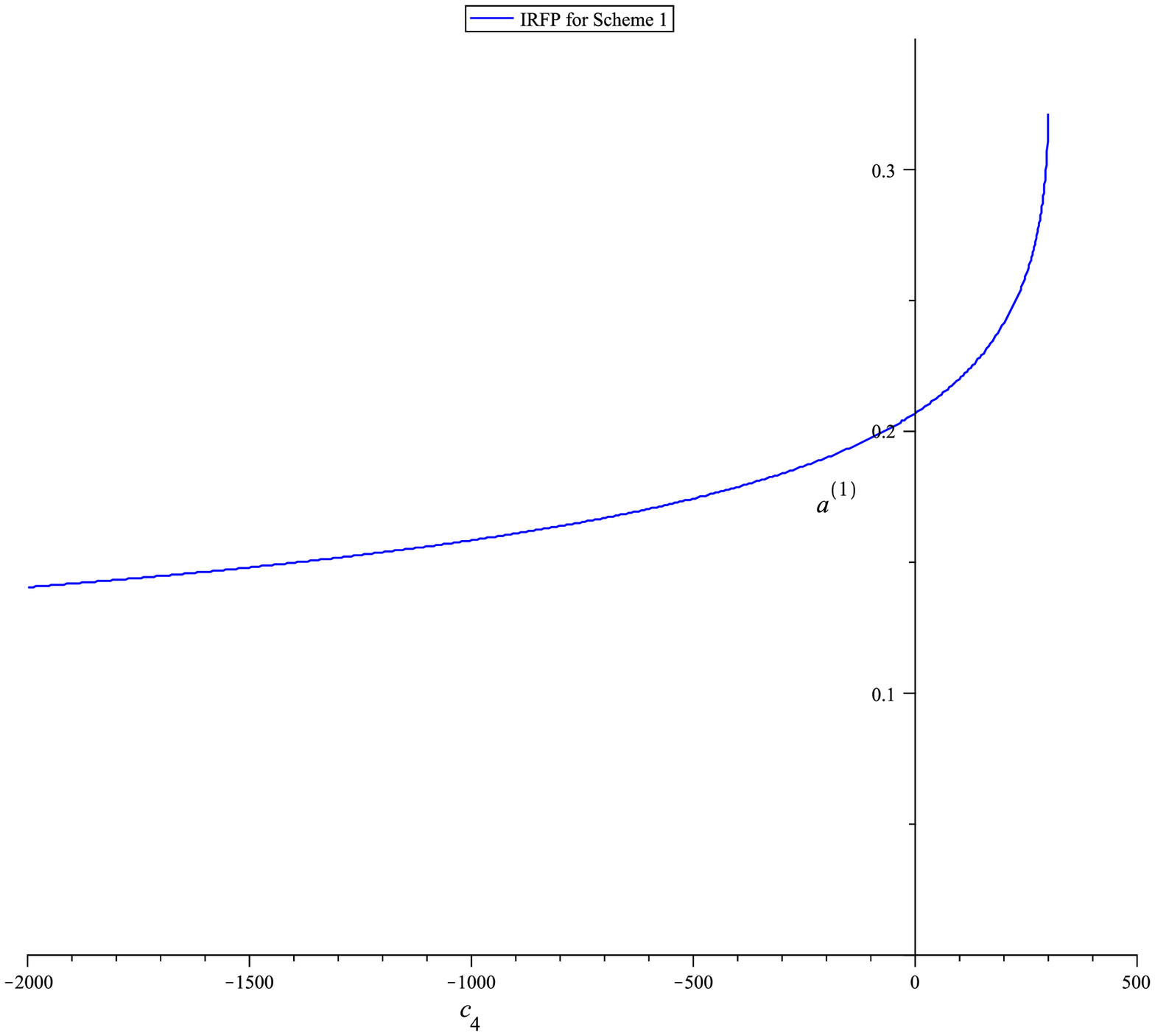}
\caption{Infrared Fixed Points (IRFPs) of the scheme-1 $\beta_{(1)}$-function at five loop order versus $c_4$ }
\label{Fig. 1}
\end{center}
\end{figure}

\end{document}